\documentclass[journal]{IEEEtran}

\usepackage{cite}
\usepackage{amsmath,amssymb,amsfonts}
\usepackage{graphicx}
\usepackage{textcomp}
\usepackage{xcolor}
\def\BibTeX{{\rm B\kern-.05em{\sc i\kern-.025em b}\kern-.08em
    T\kern-.1667em\lower.7ex\hbox{E}\kern-.125emX}}

\newcommand{\qcrank}{\textbf{QCrank}}

\newcommand{\nisq}{\textbf{NISQ}\xspace}

\newcommand{\qpd}{\textbf{QPD}\xspace}
\newcommand{\qft}{\textbf{QFT}\xspace}
\newcommand{\qaoa}{\textbf{QAOA}\xspace}
\newcommand{\sq}{\textbf{shardQ}\xspace}

\newcommand{\myvec}[1]{\vec #1}                 %
\newcommand{\dpt}{x}						    %
\newcommand{\data}{\myvec{\dpt}}			    %

\newcommand{\cnot}{CX}

\newcommand{\UCR}[1][]{
    \ifthenelse{ \equal {#1} {} }
        {\text{UCR}}
        {\text{UCR}^{(#1)}}
}

\newcommand{\UCRy}[1][]{\UCR[#1]_{y}}

\newcommand{\psifrqi}[1][]{
    \ifthenelse{ \equal {#1} {} }
    {\ket{\psi_{\text{FRQI}}}}
    {\ket{\psi_{\text{FRQI}}(#1)}}
}
\newcommand{\psiqcrank}[1][]{
    \ifthenelse{ \equal {#1} {} }
    {\ket{\psi_{\text{qcrank}}}}
    {\ket{\psi_{\text{qcrank}}(#1)}}
}
\usepackage{enumitem}
\usepackage{comment}
\usepackage{subcaption}
\usepackage[most]{tcolorbox}

\newtcolorbox{mybox}{
    colback=gray!10,       %
    colframe=white,        %
    sharp corners,         %
    boxrule=0pt,           %
    left=5pt,              %
    right=5pt,             %
    top=5pt,               %
    bottom=5pt             %
}

\usepackage{tikz}
\usetikzlibrary{quantikz2}
\usetikzlibrary{calc}
\usetikzlibrary{shapes.geometric}
\usetikzlibrary{shapes,arrows,positioning}
\usetikzlibrary{decorations.pathreplacing,math,shapes.misc}
\usepackage{amsmath,amsfonts,amsthm,mathtools,nicefrac}
\usepackage{xspace}
\usepackage{balance}
\usepackage[utf8]{inputenc}
\usepackage{caption}
\usepackage{subcaption}
\usepackage{numprint}
\npthousandsep{,}

\newtheorem{definition}{Definition}

\newtheorem{lemma}{Lemma}
\usepackage{algorithm}
\usepackage[noend]{algpseudocode}
\newcommand{\pluseq}{\mathrel{+}\!=}

\usepackage{tabularx,multirow,multicol}
\usepackage{booktabs}
\newcolumntype{L}{>{\raggedright\arraybackslash}X}
\newcolumntype{C}{>{\centering\arraybackslash}X}
\newcolumntype{R}{>{\raggedleft\arraybackslash}X}

\usepackage{environ}

\newcommand{\Ry}{R_y}
\newcommand{\Rz}{R_z}

\newcommand{\CRY}{\text{CRY}}

\newcommand{\CZ}{\text{CZ}}

\usepackage[hidelinks]{hyperref}
\usepackage[capitalize,nameinlink]{cleveref}
\Crefname{figure}{Fig.}{Figs.}
\Crefname{tabular}{Tab.}{Tabs.}
\Crefname{section}{\S}{\S}
\Crefname{theorem}{Thm.}{Thms.}
\Crefname{lemma}{Lem.}{Lems.}
\Crefname{corollary}{Cor.}{Cors.}
\Crefname{algorithm}{Alg.}{Algs.}
\Crefname{example}{Ex.}{Exs.}
\Crefname{definition}{Def.}{Defs.}

\begin{document}

\title{ShardQ: Hardware-Aware Circuit Cutting and Recomposition for Quantum Data Encoding}

\author{
  Ziqing~Guo,~
  Jan~Balewski,~
  Steven~Rayan,~
  Kewen~Xiao,~
  and~Ziwen~Pan
  \thanks{Ziqing Guo is with Texas Tech University, Lubbock, TX 79409, USA, and also with Lawrence Berkeley National Laboratory, Berkeley, CA, USA (e-mail: ziqguo@ttu.edu).}%
  \thanks{Jan Balewski is with Lawrence Berkeley National Laboratory, Berkeley, CA, USA (e-mail: balewski@lbl.gov).}%
  \thanks{Steven Rayan is with the University of Saskatchewan, Saskatoon, SK, Canada (e-mail: steven.rayan@usask.ca).}%
  \thanks{Kewen Xiao is with the Rochester Institute of Technology, Henrietta, NY, USA (e-mail: kewenxiao@gmail.com).}%
  \thanks{Ziwen Pan is with Texas Tech University, Lubbock, TX, USA (e-mail: ziwen.pan@ttu.edu).}%
}

\markboth{Accepted for publication in IEEE Transactions on Quantum Engineering}%
{Guo \MakeLowercase{\textit{et al.}}: ShardQ: Hardware-Aware Circuit Cutting and Recomposition for Quantum Data Encoding}

\maketitle

\begin{abstract}
Quantum data encoding (QDE) loads classical data into quantum states and is a core primitive for quantum machine learning and quantum algorithms, yet executing QDE circuits on connectivity-limited noisy quantum processing units (QPUs) requires a compilation-and-reconstruction pipeline that partitions the circuit across the hardware and recombines the fragment outputs. Circuit cutting and knitting provide such a route, but the benefit depends on selecting a small number of high-value cuts because quasi-probability decomposition (QPD) incurs a sampling overhead that grows as $9^k$ for $k$ cut two-qubit gates. In this study, we introduce \sq, a hardware-aware cutting and recomposition framework for structured QDE circuits with an address register, a data register, and an enumerable set of cross-register two-qubit gates under a chosen hardware layout. Its cut selector, \textbf{SparseCut}, ranks these candidate gates by hardware graph distance and selects a user-bounded cut set in $O(|\mathcal{G}_{\mathrm{cross}}|\log |\mathcal{G}_{\mathrm{cross}}|)$ time, avoiding mixed-integer-programming search in this restricted setting. The remaining truncated circuits are simulated with bond-dimension matrix product states (MPS), while the removed interactions are reconstructed using QPD on classical machines. Hardware measurements are the IBM Heron executions in \cref{tab:noise}. Cut-finding, MPS, and overhead studies up to 150+ qubit layouts are classical simulations or projections, and each such panel is labeled accordingly.
\end{abstract}


\IEEEpeerreviewmaketitle

\section{Introduction}
\label{sec:intro}

Scaling quantum computation to the utility regime, where quantum processors deliver consistent application-level advantages, remains a central challenge in the noisy intermediate-scale quantum (\nisq) era \cite{preskill2018quantum, divincenzo2025thirty, preskill2023quantum}. Despite progress in quantum algorithms \cite{montanaro2016quantum, PhysRevLett.103.150502, qaoa} and hardware \cite{de2021materials, jiang2025advancements, aghaee2025scaling}, limited qubit connectivity \cite{yuan2025full, kim2025effectiveness}, high two-qubit gate error rates \cite{hothem2025measuring}, and long-distance entanglement overhead \cite{saha2025high, ciobanu2024optimal} prevent large circuits from executing directly on current superconducting quantum processing units (QPUs). Bridging this gap is fundamentally a systems problem: it requires a compilation and execution pipeline that partitions a circuit to respect hardware constraints, executes or simulates the resulting fragments across quantum and classical resources, and recombines their outputs into the target result.

\begin{figure}[htbp]
    \centering
    \includegraphics[width=0.95\linewidth]{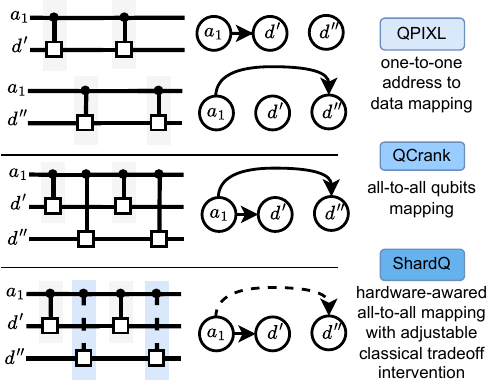}
    \caption{Overview of the proposed method compared with conventional quantum data encoding techniques.}
    \label{fig:intro}
\end{figure}

Quantum data encoding (QDE) loads classical data into quantum states and underpins resource-intensive algorithms such as Grover search \cite{grover}, Shor factoring \cite{shor1999polynomial}, and the Quantum Fourier Transform (\qft) \cite{coppersmith2002approximate}, as well as applications in cryptography \cite{pirandola2017fundamental}, optimization \cite{abbas2024challenges}, and scientific simulation \cite{bhaskar2015quantum}. In particular, quantum block data encoding \cite{camps2022fable} provides a basic primitive for singular value transformation \cite{gilyen2019quantum}, quantum walks \cite{childs2009universal}, and quantum machine learning \cite{biamonte2017quantum}. Leading encoders such as \textbf{QPIXL} \cite{QPIXL} and \qcrank \cite{jan1} (shown in \cref{fig:intro}) organize their logical qubits into an address register, a data register, and a set of cross-register entangling gates \cite{ewintang}. This regular structure exposes a well-defined and enumerable set of long-range interactions, making QDE circuits a natural compilation target for partition-and-recombine execution.

Circuit cutting and knitting realize this execution model through quasi-probability decomposition (\qpd) \cite{bravyi2016trading, pashayan2015estimating}, which replaces selected entangling gates with locally executable operations combined by classical postprocessing under local operations and classical communication (LOCC) \cite{schmitt2025cutting, brenner2023optimal}. This capability is central to hybrid high-performance-computing (\textbf{HPC})-integrated quantum platforms \cite{osti_2504206}, in which quantum resources are tightly coupled to classical accelerators for fragment reconstruction. The benefit, however, is bounded by cost. For $k$ cut CX gates, the sampling overhead is $9^k$ \cite{mitarai2021constructing,peng2020simulating,pashayan2015estimating,brenner2023optimal,schmitt2025cutting}. The number of signed terms, the number of executed measurement settings, and the variance at a fixed shot budget are separate quantities and are stated with the reconstruction in \cref{sec:globalreco,app:qpd}. Therefore, a practical pipeline must identify a small number of cuts whose hardware benefit outweighs the classical reconstruction cost.

Space-cut \cite{mitarai2021constructing} and time-cut \cite{peng2020simulating} techniques are compatible with universal two-qubit gate architectures \cite{divincenzo1995two} but are not driven by device connectivity, while hardware-aware cutting \cite{hardware4}, variational quantum clustering \cite{bermejo2023variational}, and quantum embedding analysis \cite{rath2024quantum} either rely on iterative quantum-classical optimization or are not designed for direct classical data encoding. Two concrete challenges remain. First, after layout selection and transpilation, sparse superconducting topologies impose large hardware graph distances between logical qubits, so long-range address-data interactions require routing operations that increase two-qubit error exposure. Second, cutting and knitting require both an explicit rule for selecting which entangling gates to cut and a reconstruction method that recombines the resulting fragment distributions under a controlled overhead budget.

We address these challenges with \sq, an end-to-end partition-to-recomposition pipeline for structured QDE circuits on connectivity-limited superconducting QPUs in the \nisq\ era. \sq\ targets encoders whose logical qubits are partitioned into address and data registers. Under a chosen hardware layout, its cut selector \textbf{SparseCut} enumerates the two-qubit gates crossing the register boundary and ranks them by coupling-graph distance, selecting a user-bounded cut set in $O(|\mathcal{G}_{\mathrm{cross}}|\log |\mathcal{G}_{\mathrm{cross}}|)$ time. The calibration-weighted score in \cref{eq:cal-score} is a separate comparator. The truncated circuits are then simulated with bounded-bond-dimension matrix product states (MPS), the removed interactions are reconstructed via QPD on classical accelerators, and the pipeline recomposes the encoded result with reduced long-range entangling operations. Although this method is tuned to superconducting architectures, the same candidate-set formulation can be adapted to other platforms with sparse connectivity. To the best of our knowledge, this study makes the following contributions:
\begin{itemize}
    \item We introduce \textbf{SparseCut}, a hardware-aware cut-selection rule for address-data QDE circuits. Given the cross-register candidate set $\mathcal{G}_{\mathrm{cross}}$, SparseCut selects the largest-distance candidates in $O(|\mathcal{G}_{\mathrm{cross}}|\log |\mathcal{G}_{\mathrm{cross}}|)$ time without solving a mixed-integer program.
    \item We show that, for the \qcrank\ encoders studied here, the truncated circuits admit an MPS whose exact bond dimension equals $2^{n_a}$. That bond dimension grows with the address register, and $\chi \le 128$ covers the reported window $n_a \le 7$.
    \item We reconstruct the global distribution as the signed QPD sum over the CX completions in \cref{app:qpd}. The histogram passed to the decoder clips negative quasi-counts at zero, rounds to integers, and renormalizes, as specified in \cref{eq:clip,eq:bias}.
    \item We evaluate \sq\ on IBM Heron-class hardware in \cref{tab:noise}, against an uncut \qcrank\ deployment. Cut-finding times are classical measurements. MPS curves are classical simulations, and every curve extended past the contracted sizes is marked as a projection.
    \item We validate practical utility on grayscale quantum image encoding benchmarks and report the operating range in which the hardware benefit of cutting compensates for the QPD reconstruction overhead.
\end{itemize}

The remainder of this paper is organized as follows. \cref{sec:related} reviews related work and motivates our cutting-and-knitting approach. \cref{sec:methods} details the \sq\ architecture and the algorithms for partitioning and recomposition. \cref{sec:res} reports experimental results on quantum hardware. \cref{sec:disc} discusses limitations and concludes the paper. The theoretical foundations and proofs are provided in the appendix.

\section{Related Work and Motivation}\label{sec:related}

\subsection{Circuit Cutting and Knitting Frameworks}
The concept underlying our approach traces back to the Clifford-gate group simulation protocol \cite{bravyi2005universal}, which reduces the classical-quantum computation overhead and strengthens the hybrid paradigm \cite{bravyi2016trading}. This foundation has driven both empirical \cite{peng2020simulating} and theoretical \cite{mitarai2019methodology} advances in clustered simulations of molecular systems, which were later enhanced by maximum likelihood fragment tomography \cite{perlin2021quantum}.

Existing circuit-cutting systems provide different components for partitioning, decomposition, and execution. \textbf{CutQC} \cite{tang2022cutting} provides a mixed-integer-programming (MIP) cut-selection component for small-circuit partitioning, which leads to high-quality partitions at combinatorial search cost. \textbf{Qiskit Circuit Knitting} \cite{shehzad2024automated} provides QPD decomposition, sampling, reconstruction, and MIP-based cut-finding components, which leads to automated cutting workflows subject to subcircuit and device constraints. \textbf{Qdislib} \cite{tejedor2025distributed} provides distributed HPC-QPU execution support for cut circuits, which leads to scalable orchestration while leaving topology-specific cut ranking outside its scope.
CutQC and Qiskit Circuit Knitting search general partitions and subcircuit-size constraints. \sq\ sorts the enumerable cross-register set of a structured QDE circuit by hardware distance, at cost $O(|\mathcal{G}_{\mathrm{cross}}|\log |\mathcal{G}_{\mathrm{cross}}|)$ under the sampling overhead of \cref{eq:overhead}. The timeout comparison in \cref{sec:comparative} is a search-time result for that candidate set. Fixed-$k$ routing quality is \cref{tab:routing}.

In QDE encoders such as \qcrank\ and \textbf{QPIXL}, classical values are loaded through controlled rotation layers connecting index qubits to value qubits. These layers provide an enumerable set of long-range two-qubit interactions, which leads to a cut-candidate set that can be aligned with the physical topology. Removing selected controlled rotations leaves a Clifford-dominated backbone of Hadamard and CNOT operations with local rotations, enabling bounded-bond-dimension MPS simulation before QPD reconstruction restores the cut interactions. This component structure defines the boundary between generic cut search and the structured cut rule used by \sq{}.

\subsection{Hardware-Aware Optimization}
Randomized measurement cutting for \qaoa\ \cite{lowe2023fast} provides heuristic measurement reductions, which improves sampling efficiency for variational workloads. Gradient-based reconstruction \cite{hart2024reconstructing} provides an optimization component for recovering cut observables with high fidelity. Hardware-aware compilation frameworks \cite{hardware4} provide topology-sensitive circuit transformations, which reduce routing overhead during compilation. These components act on measurement design, reconstruction, or compilation; SparseCut acts before execution by selecting long-distance cross-register gates from the QDE structure itself.

HPC-QPU integration work provides the classical capacity needed for fragment post-processing and reconstruction at scale \cite{carrera2024combining}. Classical-quantum interconnect efforts such as \textsc{NVQLink}\footnote{https://www.nvidia.com/en-us/solutions/quantum-computing/nvqlink/} provide a systems path for reducing data-movement overhead between QPUs and accelerated classical resources.

\section{Background}
\label{sec:back}
In this section, we present the essential background for the proposed \sq\ protocol. We first introduce the tensor-network simulation technique. Subsequently, we outline an approximate quantum data encoding approach. Finally, we provide the fundamental concepts underlying circuit cutting and knitting processes.

\subsection{Quantum circuit simulation}
\paragraph{MPS compilation}
The quantum state can be written as
\begin{equation}
    |\psi\rangle = \sum_{i=0}^{2^n - 1} c_i |i\rangle,
    \label{eq:basis}
\end{equation}
where classical hardware requires exponentially growing random access memory (RAM) to simulate the state vector, specifically $2^n$ complex amplitudes for $n$ qubits. Matrix product states (MPS), based on tensor networks (TN), are widely used to mitigate random access memory (RAM) overhead. For example, the Greenberger–Horne–Zeilinger (GHZ) state can be represented as
\begin{equation}
    \mathcal{G}_{s_1 s_2 s_3} = [A^{s_1} A^{s_2} A^{s_3}],
    \label{eq:ghz_mps_3}
\end{equation}
where $\mathcal{G}_{s_1 s_2 s_3}$ yields the computational basis amplitude for each $(s_1, s_2, s_3)\in\{0,1\}^3$. The contraction of a tensor network representing a quantum circuit is given by
\begin{equation}
\mathcal{A} = \sum_{{i_k}} \prod_{j} T^{[j]}_{i_{j-1}, i_j},
\end{equation}
where $T^{[j]}$ is the local tensor at site $j$, and indices $\{i_k\}$ are contracted according to the network (see Diagram~\eqref{eq:tensors}). Here, the contraction of the three tensors produces an effective tensor by summing the internal indices.
\begin{equation}
\begin{tikzpicture}[thick,scale=1.2, every node/.style={scale=1.2}]
    \draw[line width=1.5pt] (0,-0.1) -- ++(0,0.8);
    \draw[line width=1.5pt] (1.2,-0.1) -- ++(0,0.8);
    \draw[line width=1.5pt] (2.4,-0.1) -- ++(0,0.8);
    \draw[line width=1.5pt] (0, 0) -- (2.4, 0);
    \node[circle, fill=red!70, minimum size=10pt, inner sep=0pt] (A1) at (0,0) {};
    \node[circle, fill=red!70, minimum size=10pt, inner sep=0pt] (A2) at (1.2,0) {};
    \node[circle, fill=red!70, minimum size=10pt, inner sep=0pt] (A3) at (2.4,0) {};
    \node[below=3pt] at (A1) {$G^{s_1}$};
    \node[below=3pt] at (A2) {$G^{s_2}$};
    \node[below=3pt] at (A3) {$G^{s_3}$};
    \draw[line width=2pt, -{Latex[length=4mm, width=2mm]}] (2.7,0) -- (3.3,0);
    \draw[rounded corners=8pt, thick, line width=1.5pt, fill=gray!20] (3.5,-0.2) rectangle (4,0.7);
    \draw[line width=1.5pt] (4,0) -- (4.125,0);
    \draw[line width=1.5pt] (4.3, -0.1) -- ++(0,0.8);
    \node[circle, line width=1.5pt, fill=red!80, minimum size=10pt, inner sep=0pt] (C) at (4.3,0) {};
\end{tikzpicture}
\label{eq:tensors}
\end{equation}
An important advantage of this method is that for circuits exhibiting structured entanglement, the time complexity is $O(\text{poly}(N)\cdot 2^{w})$~\cite{berezutskii2025tensor}, where $w$ represents the minimal width determined by the circuit connectivity.

\subsection{Gate-based quantum data encoding}
\label{bg: gate}
Data encoding is essential for embedding and image encoding, as it enables a compact Hilbert space representation of classical data. Quantum data encoding transforms classical information into quantum states using three different methods. \textit{Basis encoding} maps discrete values directly to computational basis states using binary representation; for example, \([01,11]\) becomes $\ket{x^1}=\ket{01}$ and $\ket{x^2}=\ket{11}$. \textit{Amplitude encoding} embeds normalized data $\data$ into quantum amplitudes $\sum_i \alpha_i \ket{i}$, subject to normalization. \textit{Angle encoding} maps features to qubit states using rotation gates, so each feature is encoded as $\cos(\theta_i/2)\ket{0} + \sin(\theta_i/2)\ket{1}$, with $\theta_i$ rescaled to $[0, \pi]$. The resulting quantum state is
\begin{equation}
    |\psi\rangle = \bigotimes_{i=1}^n \left( \cos\frac{\theta_i}{2} \ket{0} + \sin\frac{\theta_i}{2} \ket{1} \right).
    \label{eq:qs}
\end{equation}
Following angle encoding, \qcrank\ encodes classical data with the Unitary Controlled Rotation (\(\UCRy\)) gate
\begin{equation}
    \UCRy(\theta) =
    \begin{pmatrix}
    R_y(\theta_0) &        & \\
                & \ddots & \\
                &        & R_y(\theta_{2^{n_a}-1})
    \end{pmatrix}.
\label{eq:ucry_mat}
\end{equation}
Each address qubit configuration (representing the position of each classical data input) $|i\rangle$ is prepared in a superposition using the Walsh–Hadamard Transform (WHT). For each address, the associated data values are encoded onto the data qubits by applying single-qubit rotations. Specifically, for every address, each data qubit receives a rotation by an angle corresponding to the classical data value assigned to that address and data qubit.
\begin{equation}
    |c_{i,j}\rangle = \cos(\theta_{i,j}/2)\ket{0} + \sin(\theta_{i,j}/2)\ket{1}.
    \label{eq:encode}
\end{equation}
The \(\UCRy\) gate applies these rotations in a block-diagonal form controlled by the address qubits. Thus, the \qcrank\ encoder produces
\begin{equation}
    |\psi_{\mathrm{qcrank}}(\vec{\theta})\rangle = \frac{1}{\sqrt{2^{n_a}}} \sum_{i=0}^{2^{n_a}-1} |i\rangle \otimes |c_{i,0}\rangle \otimes \cdots \otimes |c_{i,n_d-1}\rangle.
    \label{eq:qcrank}
\end{equation}
with each $|c_{i,j}\rangle$ angle-encoded via \(\UCRy\), $n_a$ address qubits and $n_d$ data qubits. For example, encoding the 3D tensor $(1,2,1)$ yields the circuit
\begin{equation}
     \begin{quantikz}
    \lstick{$a_0$} & \gate{H}\slice{} & & \ctrl{1} & &\ctrl{1} & \meter{}  \\
    \lstick{$d_1$} & &\gate{R_y(\theta_1)} & \targ{} & \gate{R_y(\theta_2)}& \targ{}  & \meter{}
    \end{quantikz}
\label{circ:q11}
\end{equation}
resulting in $|\psi_{\text{qcrank}}\rangle = \frac{1}{\sqrt{2}} \left( |0\rangle|0\rangle + |1\rangle \left[ c|0\rangle + s|1\rangle \right] \right)$, where $c = \cos\left(\frac{\theta_1+\theta_2}{2}\right)$, $s = \sin\left(\frac{\theta_1+\theta_2}{2}\right)$. To retrieve classical data, projective measurements estimate $|c|^2$ and $|s|^2$, and the encoded angle is reconstructed by
\begin{equation}
\theta_1 + \theta_2 = 2\arctan\left(\sqrt{\frac{|s|^2}{|c|^2}}\right),
\end{equation}
assuming that the angle is rescaled to $[0, \pi]$ using \(arccos(0,1)\) via EVEN encoding \cite{balewski2025ehands}.

\subsection{Circuit Cutting and Knitting}
\label{bg:cut and knit}
The circuit knit-and-cut technique enables the efficient simulation of large quantum circuits by dividing them into smaller, manageable subcircuits $(SC)$, which are simulated separately and then recombined to construct global observables. This approach exploits the fact that the expectation value of an operator on the entire circuit, such as $\mathcal{S}(\mathbf{U}_1 \otimes \mathbf{U}_2)$~\cite{mitarai2021constructing}, can be represented as a linear combination of the expectation values of each subcircuit, each weighted by classical coefficients, where \(\mathcal{S}\) represents the operator in the Pauli groups. For two arbitrary unitaries $U$, the observable in the QPD framework is reconstructed as
\begin{equation}
    \mathcal{S}(\mathbf{U}_1 \otimes \mathbf{U}_2) = \mathbf{c}^T \mathbf{S} + \frac{1}{4}\sin(2\theta)\sin(2\phi) \sum_{\boldsymbol{\alpha}} \alpha_1 \alpha_2\, \mathcal{R}(\boldsymbol{\alpha}).
\end{equation}
Here, $\mathbf{S}$ contains the expectation values of the subcircuits with different operator insertions at the cut, and $\mathbf{c}$ provides the corresponding weights of these expectation values. The sum over $\boldsymbol{\alpha}$ captures the correlations introduced by the cut (see Appendix A in \cite{mitarai2019methodology}). This technique relies on efficient classical post-processing to recover the outcome of the original circuit (see Table 1 in~\cite{schmitt2025cutting}).
Here, we provide the general QPD sampling overhead for a CX gate for our protocol.
\begin{definition}[QPD overhead for CX gate]
Let $C$ be a quantum circuit with $k$ cut CNOT gates. The QPD sampling overhead is
\begin{equation}
    O_{\mathrm{QPD}}(C) = 3^{2k} = 9^{k},
    \label{eq:overhead}
\end{equation}
where $k$ is the number of cut CX gates. The value $9^{k}=3^{2k}$ is the sampling overhead: the factor by which the shot budget grows so that the quasiprobability estimator keeps the same variance \cite{mitarai2021constructing,peng2020simulating,pashayan2015estimating,brenner2023optimal,schmitt2025cutting}. Each cut contributes two single-qubit factors, and each factor of base 3 is the three non-trivial Pauli insertions ($X$, $Y$, $Z$) on that qubit, as drawn in \cref{fig:cut}. This overhead is distinct from the number of signed terms and from the number of executed circuits, which are given in \cref{sec:globalreco,tab: pauli,tab:phi_prime_bases}. At a fixed total of $N$ shots, the variance of the knitted estimator scales as $\mathcal{O}(9^{k}/N)$.
\end{definition}
\begin{figure}[htbp]
    \centering
    \resizebox{\columnwidth}{!}{
    \begin{tikzpicture}[thick, > = latex]

    \tikzset{scissor/.pic={
      \begin{scope}[scale=0.3]
            \draw[thick] (-0.6,-0.6) -- (0.6,0.6);
            \draw[thick] (0.6,-0.6) -- (-0.6,0.6);
            \draw[thick,fill=white] (0.4,0.4) circle (0.22);
            \draw[thick,fill=white] (-0.4,0.4) circle (0.22);
        \end{scope}
    }}
    \fill[gray!20] (-0.2,2.7) rectangle (7.2,4.3); %
    \fill[purple!10] (-0.2,0.3) rectangle (7.2,2.3); %

    \foreach \y in {1,2,3,4} {
      \draw[black] (0,\y) -- (7,\y);
    }

    \foreach \x in {0.5,2.5,4.5,6.5} {
      \draw[fill=white] (\x,3.7) rectangle +(0.6,0.6);
      \draw[fill=white] (\x,0.7) rectangle +(0.6,0.6);
    }
    \draw[fill=pink!70] (3.5,2.7) rectangle +(0.6,1.6);
    \draw[fill=pink!70] (5.5,1.7) rectangle +(0.6,1.6);
    \pic at (3.8,3.5) {scissor};
    \pic at (5.8,2.5) {scissor};
    \draw[->,line width=1.2pt] (7.8,2) -- (9.2,2);

    \fill[gray!20] (10,2.7) rectangle (17.2,4.3);
    \fill[purple!10] (10,0.3) rectangle (17.2,2.3);

    \foreach \y in {1,2,3,4} {
      \draw[black] (10,\y) -- (17,\y);
    }
    \foreach \x in {10.5,12.5,14.5,16.5} {
      \draw[fill=white] (\x,3.7) rectangle +(0.6,0.6);
      \draw[fill=white] (\x,0.7) rectangle +(0.6,0.6);
    }
    \measure{16.5}{2}

    \draw[fill=pink!70] (13.5,3.7) rectangle +(0.6,0.6);
    \draw[fill=pink!70] (13.5,2.7) rectangle +(0.6,0.6);
    \draw[fill=pink!70] (15.5,2.7) rectangle +(0.6,0.6);
    \draw[fill=pink!70] (15.5,1.7) rectangle +(0.6,0.6);

    \node[gray!60!black] at (1.5,3.5) {$\bar{A}$};
    \node[purple!60!black] at (1.5,1.5) {$\bar{B}$};
    \node[gray!60!black] at (11.5,3.5) {$\bar{A}$};
    \node[purple!60!black] at (11.5,1.5) {$\bar{B}$};
    \end{tikzpicture}
    }
    \caption{The generic example of cutting two qubit gates into separate one qubit gates. Note that, the right side only shows one of the subcircuits.}
    \label{fig:cut}
\end{figure}

\section{Methods}
\label{sec:methods}
\begin{figure}[ht]
    \centering
    \includegraphics[width=0.99\linewidth]{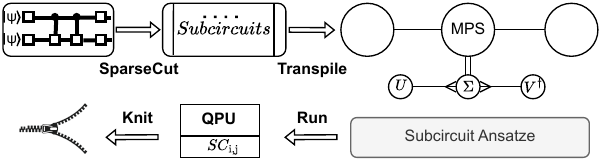}
    \caption{The \sq\ protocol: (1) The original data encoder circuit is analyzed by \textbf{SparseCut} to identify long-distance cross-partition gates for cutting. (2) The truncated circuit (with cut gates removed) is simulated using MPS, exploiting its Clifford-dominated structure for improved efficiency. (3) Each removed CX is executed as one of the measurement settings in \cref{tab:phi_prime_bases}, denoted $\mathrm{SC}_{i,j}$. (4) Results are accumulated with the QPD coefficients and parity signs, then clipped and renormalized as in \cref{eq:clip}.}
    \label{fig:method}
\end{figure}

\cref{fig:method} shows the end-to-end \sq\ protocol that leverages circuit cutting, MPS simulation of truncated circuits, and global result reconstruction via QPD. The proposed protocol targets gate-based quantum simulations on connectivity-constrained hardware. Specifically, we provide a deterministic cutting strategy, \textbf{SparseCut}, for quantum data encoders that produces truncated circuits suitable for efficient MPS simulation. In addition, we propose a generic global measured bit string reconstruction algorithm.

\begin{figure*}[ht]
    \centering
    \includegraphics[width=0.99\linewidth]{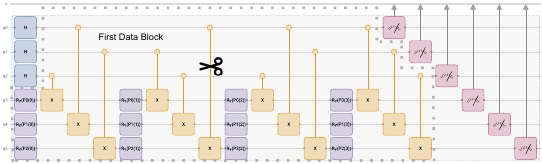}
    \caption{Logical \qcrank\ circuit before placement, with address qubits $q_0,q_1,q_2$ and data qubits $q_3,q_4,q_5$. Scissors mark the cross-register CX gates ranked by \cref{alg:sparse_cut_selection}. Only the first data block is drawn; the remaining blocks follow Fig.~1(c) of \cite{jan1}. Each $R_y$ angle is an encoded value indexed by the address. \Cref{fig:coupling} is the hardware placement of these same logical qubits.}
    \label{fig:qcrank_cut}
\end{figure*}
\begin{figure}[htbp]
    \centering
    \includegraphics[width=0.98\linewidth]{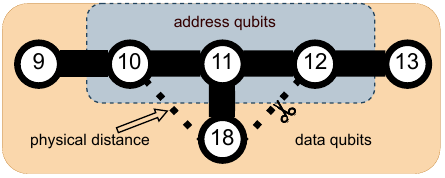}
    \caption{Placement $\phi$ of the logical qubits in \cref{fig:qcrank_cut} on the \texttt{ibm\_pittsburgh} heavy-hex coupling graph \cite{ibm_pitt}. Blue vertices are the address qubits $q_0,q_1,q_2$. Yellow vertices are the data qubits $q_3,q_4,q_5$. The dashed line joins the pair with the largest shortest-path distance $d_T(\phi(q_i),\phi(q_j))$, which is the first gate selected by \cref{alg:sparse_cut_selection}. This figure is that layout. It is the coupling graph on which \cref{fig:qcrank_cut} is scored. The level-3 preset transpiler then inserts SWAPs; a fixed-$k$ count of the resulting two-qubit gates is reported in \cref{tab:routing}.}
    \label{fig:coupling}
\end{figure}
\subsection{Cutting Algorithm}
Similar to the use of permuted controlled unitary operations in the \qft, the \qcrank\ encoder employs layers of gray-coded CX gates to facilitate data entanglement and connectivity in a high-dimensional Hilbert space. Because each cut contributes to the QPD overhead in \cref{eq:overhead}, \sq\ uses a fixed cut budget and selects only the cross-register gates that are most expensive under the hardware layout. \Cref{fig:qcrank_cut} shows an example for three address qubits and three data qubits.

\begin{definition}[SparseCut candidate set]
\label{def:sparsecut-candidates}
Let $C$ be a QDE circuit with address register $A$ and data register $D$. Let $T=(V_T,E_T)$ be the hardware coupling graph and let $\phi$ map each logical qubit to a physical qubit. The SparseCut candidate set is
\begin{equation}
\mathcal{G}_{\mathrm{cross}}(C,A,D)
=
\{g=(q_i,q_j) \in C_2:\ |\{q_i,q_j\}\cap A|=1\},
\end{equation}
where $C_2$ is the set of two-qubit gates in $C$. Each candidate is assigned the hardware score
\begin{equation}
s(g)=d_T(\phi(q_i),\phi(q_j)),
\label{eq:sparsecut-score}
\end{equation}
where $d_T$ is the shortest-path distance on the coupling graph. The ranking evaluated as SparseCut uses this distance. The calibration-weighted comparator uses a different score,
\begin{equation}
s_{\mathrm{cal}}(g)=0.35\, d_T + 0.30\, e(g) + 0.35\, \bigl(100\, \varepsilon_{\mathrm{SWAP}}\bigr),
\label{eq:cal-score}
\end{equation}
where $e(g)$ is the average CZ error of the two physical qubits divided by the device median and $\varepsilon_{\mathrm{SWAP}}=1-(1-3\epsilon_2)^{\max(d_T-1,0)}$, with $\epsilon_2$ the median CZ error. \Cref{tab:routing} compares distance ranking with logical-distance ranking and with uniform random selection. It does not apply \cref{eq:cal-score}.
\end{definition}

Here, $q_i$ and $q_j$ refer to logical qubits, and $d_T$ is the hardware graph distance used consistently throughout our methodology. First, we recall that the address and data qubits encode \cref{circ:q11}
$\text{data} = \begin{bmatrix}
    \begin{bmatrix}
        \theta_{0,0,0} \\
        \theta_{0,1,0}
    \end{bmatrix}
\end{bmatrix}.$
Here, we denote $\text{data}[c][a][d]$, where the parameters correspond to the circuit, address, and data index, respectively.
\Cref{fig:qcrank_cut} is the logical gate list. \Cref{fig:coupling} assigns those logical qubits to physical qubits on the \texttt{ibm\_pittsburgh} coupling map \cite{ibm_pitt}. The score in \cref{eq:sparsecut-score} is the shortest path on that graph.
\begin{algorithm}[t]
\caption{SparseCut Selection}
\label{alg:sparse_cut_selection}
\begin{algorithmic}[1]
    \Require Circuit $C$, address register $A$, data register $D$, layout $\phi$, coupling graph $T$, cut budget $K$
    \Ensure Cut index set $\mathcal{S}$ with $|\mathcal{S}|\le K$
    \State $\mathcal{G}_{\text{cross}} \gets \varnothing$ \Comment{Initialize candidate set}
    \For{each two-qubit gate $g=(q_i,q_j)$ in $C$}
        \If{$|\{q_i, q_j\} \cap A| = 1$} \Comment{Check boundary crossing}
            \State $d \gets d_T(\phi(q_i),\phi(q_j))$ \Comment{Hardware graph distance}
            \State $\mathcal{G}_{\text{cross}} \gets \mathcal{G}_{\text{cross}}\cup \{(\text{id}(g), d)\}$
        \EndIf
    \EndFor
    \State Sort $\mathcal{G}_{\text{cross}}$ by $d$ descending \Comment{Prioritize long-range gates}
    \State \Return Top $\min(|\mathcal{G}_{\text{cross}}|, K)$ elements
\end{algorithmic}
\end{algorithm}
By considering the address and data registers, \cref{alg:sparse_cut_selection} identifies cross-partition gates and ranks them using the hardware graph distance in \cref{eq:sparsecut-score}. The hyperparameter \texttt{max\_cuts} bounds the number of selected gates and therefore controls the QPD overhead.

\paragraph{Complexity Analysis}
We formalize the cost of SparseCut as a ranking pass over the candidate set.

\begin{lemma}
\label{lem:sparsecut-complexity}
For any QDE circuit $C$ with candidate set $\mathcal{G}_{\mathrm{cross}}$, SparseCut runs in
\begin{equation}
O(|C|+|\mathcal{G}_{\mathrm{cross}}|\log |\mathcal{G}_{\mathrm{cross}}|)
\end{equation}
time and uses $O(|\mathcal{G}_{\mathrm{cross}}|)$ auxiliary storage. If a circuit family satisfies $|\mathcal{G}_{\mathrm{cross}}|=O(n)$ for $n$ logical qubits under the chosen encoding and layout regime, then the cut-selection step is $O(|C|+n\log n)$.
\end{lemma}

\begin{proof}
The algorithm first scans the gate sequence of $C$ once and tests whether each two-qubit gate crosses the address-data register boundary. This enumeration step is linear in the circuit size and therefore costs $O(|C|)$. Each identified candidate gate is then assigned the hardware graph-distance score $d_T(\phi(q_i),\phi(q_j))$ defined in \cref{alg:sparse_cut_selection}. Computing these scores is linear in $|\mathcal{G}_{\mathrm{cross}}|$ once the layout and coupling graph are fixed, while sorting the scored candidates dominates this stage with cost $O(|\mathcal{G}_{\mathrm{cross}}|\log |\mathcal{G}_{\mathrm{cross}}|)$. The final selection returns the first $K$ candidates from the sorted list, adding $O(K)$ time and no asymptotically larger storage. Because $K\le |\mathcal{G}_{\mathrm{cross}}|$ and is fixed by the QPD cut budget in our experiments, this term is absorbed by the enumeration and sorting costs. Summing these terms gives $O(|C|+|\mathcal{G}_{\mathrm{cross}}|\log |\mathcal{G}_{\mathrm{cross}}|)$ time and $O(|\mathcal{G}_{\mathrm{cross}}|)$ auxiliary storage for the candidate list. If $|\mathcal{G}_{\mathrm{cross}}|=O(n)$ for the considered circuit family, the bound becomes $O(|C|+n\log n)$.
\end{proof}
This result should be interpreted as the cost of a deterministic hardware-ranking rule, not as an optimal cut-count solver. In expanded \qcrank\ circuits, the candidate-set bound $|\mathcal{G}_{\mathrm{cross}}|$ is the relevant quantity when the address-register size is varied; the linear corollary applies to regimes where the number of cross-register candidates grows linearly with the number of logical qubits.

\subsection{MPS Simulation of Truncated Circuits}
\label{sec:mps_clifford}
A key insight of \sq\ is that after applying SparseCut, the resulting subcircuits are dominated by Clifford gates (Hadamard, CNOT, Pauli gates), enabling efficient classical simulation via Matrix Product States (MPS). This property fundamentally distinguishes our approach from generic circuit cutting, where the post-cut subcircuits may remain intractable.

\paragraph{Circuit Truncation}
The complete encoder circuit $C$ is partitioned by SparseCut into a set of cut gates $\mathcal{G}_{\mathrm{cross}}^{\star}$ and a truncated circuit $C^{\mathrm{trunc}}$. All two-qubit operations that couple the address register $A$ to the data register $D$ are ranked according to the hardware graph distance $d_T$, and the $k=\texttt{max\_cuts}$ gates with the largest score are selected for cutting the circuit. Formally:
\begin{equation}
    C^{\mathrm{trunc}} = C \setminus \mathcal{G}_{\mathrm{cross}}^{\star},
\end{equation}
where the cut gates are removed from the circuit and handled via QPD reconstruction (\cref{sec:globalreco}).

\paragraph{MPS tractability}
The truncated circuit $C^{\mathrm{trunc}}$ enables the removal of high-scoring long-range address-data gates, namely those with the largest hardware graph-distance score $d_T$ under the chosen layout (\cref{eq:sparsecut-score}), which reduces entanglement across the partition. Specifically, for \qcrank-style encoders, the resulting circuit architecture is restricted to three localized components. First, the address register is limited to Hadamard and short-range CNOT gates, maintaining a Clifford-dominant structure. Second, the data register consists primarily of local $R_y(\theta)$ rotations, which, while non-Clifford, remain computationally inexpensive single-qubit operations because they act locally on the tensor structure without increasing the bond dimension. Finally, the remaining cross-partition interactions have lower hardware graph-distance scores than the removed gates, reducing the routing burden.
For the \qcrank\ encoders in \cref{sec:res}, the exact bond dimension of this truncated state is $\chi = 2^{n_a}$. The window $\chi \le 128$ is the range $n_a \le 7$ reported in \cref{fig:mps_bonddim}. Each added address qubit doubles $\chi$. The same equality is a property of this encoder family.

\paragraph{Scalability Analysis}
The computational complexity of the MPS simulation is characterized by $O(n^2 \cdot \chi^3)$ for $n$ qubits and bond dimension $\chi$ \cite{vidal2003efficient}. When SparseCut yields fragments with bounded $\chi$, the resulting simulation is polynomial in $n$ for the targeted structured circuits. This contrasts with the $O(2^n)$ memory growth of statevector simulation. \Cref{fig:mps_scaling} in \cref{sec:res} reports MPS scaling and projected large-$n$ behavior up to 150 qubits. Unlike general-purpose Approximate Quantum Compilation (AQC) techniques \cite{robertson2025approximate}, which optimize parameterized ansatzes to approximate arbitrary target states, \sq\ uses the structure of QDE circuits to obtain MPS-tractable truncated circuits. The removed gates are then reconstructed through QPD rather than variationally approximated.

\subsection{Global Reconstruction}
\label{sec:globalreco}
We provide the pseudocode in \cref{alg:recon} for global bit-string reconstruction. To recover the original circuit statistics, we start with the basic \qpd\ recursion. For a single cut, the quasi-probability expansion of an
observable \(\mathcal{M}\) acting on circuit \(\mathcal{C}\) is
\begin{equation}
T(\mathcal{C},\mathcal{M})=
\sum_{i=1}^{8} c_i\,
      T(\mathcal{C}',\mathcal{M}_i),
\end{equation}
where
\(T(\mathcal{C},\mathcal{M})
  =\operatorname{Tr}[\mathcal{M}\,\mathcal{C}(\rho)]\)
and \(\mathcal{C}'\) denotes the circuit after inserting one Pauli
completion.
\begin{algorithm}[tb]
\caption{Reconstruct global counts}
\label{alg:recon}
\begin{algorithmic}[1]
\Require Result records $\mathcal R$; coefficient map $c(\cdot)$; observed-bit count $n_o$
\Ensure Global counter $\mathcal G: \Sigma^* \to \mathbb{R}$
\State $\mathcal G \gets [s \mapsto 0]$
\ForAll{$r \in \mathcal R$}
    \State $\mathsf{cnt} \gets \textsc{Histogram}(r)$
    \ForAll{$(s,n)\in \mathsf{cnt}$}
        \State $\text{obs}\gets s[1:n_o]$; \quad $\text{qpd}\gets s[n_o+1:]$
        \If{$|\text{qpd}|>0$}
            \State $\sigma \gets \textsc{Parity}(\text{qpd})$
        \Else
            \State $\sigma \gets 1$
        \EndIf
        \State $\mathcal G(\textsc{Reverse}(\text{obs})) \pluseq c(r)\cdot \sigma \cdot n$
    \EndFor
\EndFor
\State \Return $\{s \mapsto \max(0,\lfloor v+0.5\rfloor) \mid (s,v)\in \mathcal G\}$
\end{algorithmic}
\end{algorithm}
The accumulator $\mathcal{G}(s)$ stores a signed quasi-count. The return line replaces each value $v$ by $\max(0,\mathrm{round}(v))$, which drops every negative quasi-count and rounds the remainder to an integer. \Cref{eq:rvf} and the RMSE are computed from the renormalized distribution
\begin{equation}
p_s = \frac{v_s^{+}}{\sum_t v_t^{+}}, \qquad v_s^{+}=\max(v_s,0).
\label{eq:clip}
\end{equation}
The signed sum below is the quasiprobability estimator. Clipping and renormalization are a subsequent classical map. Their bias relative to that estimator is summarized by the negative-mass fraction and the total variation
\begin{equation}
\eta = \frac{\sum_s \max(-v_s,0)}{\sum_s |v_s|}, \qquad
\mathrm{TV}(p,\tilde p) = \frac12 \sum_s \bigl|p_s - \tilde p_s\bigr|,
\label{eq:bias}
\end{equation}
where $\tilde p$ is the signed accumulator divided by its total sum. \Cref{tab:noise} reports RVF after \cref{eq:clip}. The stored hardware histograms are those post-clip counts, so $\eta$ is not listed as a column of \cref{tab:noise}.
By iterating the rule over \(M\) cuts gives
\begin{equation}
T(\mathcal{C},\mathcal{M})=
\sum_{\boldsymbol{\alpha}\in\{1,\dots,8\}^{M}}
   \Bigl(\prod_{m=1}^{M}c_{\alpha_m}\Bigr)\,
   T\bigl(\mathcal{C}',\mathcal{M}_{\boldsymbol{\alpha}}\bigr),
   \label{eq: T1}
\end{equation}
with \(\boldsymbol{\alpha}=(\alpha_1,\dots,\alpha_M)\) enumerating the
\(8^{M}\) signed completions in \cref{tab: pauli}. For one CX, the operator identity in \cref{eq: CNOT_all} has four Pauli terms, \cref{tab: pauli} lists eight signed completions, and \cref{tab:phi_prime_bases} realizes those eight completions as six executed measurement settings. The sampling overhead remains $9^{k}$ in \cref{eq:overhead} \cite{mitarai2021constructing,peng2020simulating,pashayan2015estimating,brenner2023optimal,schmitt2025cutting}, and the variance at a fixed total of $N$ shots remains $\mathcal{O}(9^{k}/N)$.
Removing every cut edge splits \(\mathcal{C}'\) into \(K\) independent
fragments,
\begin{equation}
    \mathcal{C}'=C^{(1)}\sqcup C^{(2)}\sqcup\cdots\sqcup C^{(K)}.
    \label{eq: Cp}
\end{equation}
Because of the independent expectation values, we provide
\begin{equation}
T\!\bigl(\mathcal{C}',\mathcal{M}_{\boldsymbol{\alpha}}\bigr)=
\prod_{k=1}^{K}
T\Bigl(
     C^{(k)},
     \mathcal{M}^{(k)}_{\boldsymbol{\alpha}_{\mathrm{cuts}(k)}}
   \Bigr),
\label{eq:subcircs}
\end{equation} from \cref{eq: T1} and \cref{eq: Cp}
where
\(\boldsymbol{\alpha}_{\mathrm{cuts}(k)}\subset\boldsymbol{\alpha}\)
holds only the remaining components that act on the fragment \(k\).
In \cref{fig:method}, $\mathrm{SC}_{i,j}$ is fragment $C^{(i)}$ under one distinct measurement setting. For the CX decomposition in \cref{tab:phi_prime_bases}, that is six settings per local cut. Executing these settings on the QPU provides the conditional
probabilities
\(P^{(i)}\bigl(b_i \mid \boldsymbol{\alpha}_{\mathrm{cuts}(i)}\bigr)\).
Inserting these probabilities into the squared-modulus of
\eqref{eq:subcircs} yields the knitted distribution
\begin{equation}
P(\mathbf{b}) =
\sum_{\boldsymbol{\alpha}}
  \Bigl(\prod_{m=1}^{M} c_{\alpha_m}\Bigr)
  \prod_{i=1}^{K}
  P^{(i)}\bigl(
    b_i \mid \boldsymbol{\alpha}_{\mathrm{cuts}(i)}
  \bigr).
\end{equation}

For the CX cuts in this work, each cut multiplies the sampling overhead by 9 \cite{mitarai2021constructing,peng2020simulating,pashayan2015estimating,brenner2023optimal,schmitt2025cutting}. For a cut set $S$, holding the estimator variance fixed therefore costs a factor $9^{|S|}$. When the fragments have the bond dimension in \cref{fig:mps_bonddim}, the classical arithmetic on top of that shot factor is polynomial in $n$.
\Cref{alg:recon} multiplies each local count by its QPD coefficient and parity sign and adds it into the signed accumulator $\mathcal{G}(s)$. The probabilities passed to RVF and RMSE are those of \cref{eq:clip}.

\section{Evaluation}
\label{sec:res}

We organize the evaluation around four research questions that isolate the cost and the benefit of each stage of the \sq\ pipeline:
\begin{itemize}[leftmargin=*,nosep]
\item \textbf{RQ1 (Cut-selection cost).} On address--data QDE circuits, how does SparseCut's sort over $\mathcal{G}_{\mathrm{cross}}$ compare in wall-clock search time with MIP cut finders under a 60-second timeout? (\cref{sec:comparative})
\item \textbf{RQ2 (Hardware benefit).} On real superconducting hardware, does cutting the highest graph-distance cross-register gates reduce end-to-end reconstruction error relative to an uncut deployment? (\cref{sec:comparative}, \cref{tab:noise})
\item \textbf{RQ3 (Classical tractability).} For these \qcrank\ encoders, does the MPS bond dimension stay at $2^{n_a}$, and thus at or below 128 for $n_a \le 7$, as the data register grows? (\cref{sec:scalability})
\item \textbf{RQ4 (Operating regime).} Over what cut budget $k$ does the routing-error reduction from cutting outweigh the $9^k$ QPD reconstruction overhead? (\cref{sec:costmodel}, \cref{sec:scalability})
\end{itemize}

\subsection{Experimental Setup}
\label{sec:setup}

\paragraph{Hardware}
We evaluate on IBM Heron processors: \texttt{ibm\_pittsburgh} (Heron-r3, 156 qubits, median CZ error $1.46 \times 10^{-3}$) and \texttt{ibm\_marrakesh} (Heron-r2, median CZ error $2.1 \times 10^{-3}$). The median CZ error is the backend calibration value recorded with each job. Each submitted circuit, including each QPD configuration, receives $10^6$ shots, so the aggregate shot count is $10^6$ times the number of executed configurations. Transpilation uses the Qiskit preset pass manager at optimization level 3 with transpiler seed 42. Gate and measurement twirling use 60 randomizations, and dynamical decoupling is off. These are the settings of the submission configuration \cite{shardq}. \Cref{tab:noise} reports one execution per column. Cut-finding times and MPS timings are classical runs. To accelerate classical subcircuit recombination, we utilized the GPU nodes of the Perlmutter supercomputer, each equipped with four 80 GB NVIDIA A100 GPUs and a single AMD Milan CPU. The post-processing workflow is containerized using Podman to enable efficient, scalable execution across the distributed architecture \footnote{https://docs.nersc.gov/systems/perlmutter/running-jobs/}.

\paragraph{Implementation}
\sq\ is implemented in Python using Qiskit 2.0 for circuit construction, Qiskit Aer's MPS simulator (with bounded bond dimension $\chi \le 128$) for scalable simulation, and our proposed QPD global reconstruction routines \cref{alg:recon}. CutQC and Qiskit Circuit Knitting baselines use their default configurations with MIP solver timeout set to 60 seconds.

\paragraph{Metrics}
Let $p$ denote the reconstructed bit-string distribution and $q$ denote the reference distribution obtained from noiseless simulation or the uncut comparison specified in each experiment. We report reconstruction vector fidelity (RVF) as the squared Bhattacharyya fidelity
\begin{equation}
\mathrm{RVF}(p,q)=\left(\sum_{b\in\{0,1\}^{n}}\sqrt{p_b q_b}\right)^2,
\label{eq:rvf}
\end{equation}
where larger values indicate closer agreement. For decoded image data, we report
\begin{equation}
\mathrm{RMSE}=\sqrt{\frac{1}{N}\sum_{i=1}^{N}(\hat{x}_i-x_i)^2},
\end{equation}
where $x_i$ and $\hat{x}_i$ are the reference and reconstructed pixel values. RVF compares bit-string distributions through \cref{eq:rvf}. RMSE is computed after the decoder maps those probabilities to pixel angles by $\theta = 2\arctan\sqrt{|s|^2/|c|^2}$, as in \cref{sec:back}. A timing error bar is the standard deviation over the repeats named in that caption. When $R>1$ independent hardware jobs are available, the interval is $\bar{x} \pm 1.96\, s/\sqrt{R}$. \Cref{tab:noise} lists one execution per column, so those entries are single points.

\subsection{Comparative Analysis}
\label{sec:comparative}

We evaluate \sq\ against three baselines:
\begin{enumerate}[leftmargin=*,nosep]
    \item \textbf{Uncut}: Standard Qiskit transpilation (optimization level 3) without circuit cutting, representing the current deployment approach.
    \item \textbf{CutQC} \cite{tang2022cutting}: State-of-the-art MIP-based circuit cutting framework optimizing for minimum cut count.
    \item \textbf{Qiskit Circuit Knitting} \cite{shehzad2024automated}: IBM's official circuit decomposition toolbox with automated cut finding.
\end{enumerate}
Here, we note that the constructed circuit follows the diagram shown in \cref{fig:qcrank_cut}.

\begin{figure}[htbp]
    \centering
    \includegraphics[width=0.98\linewidth]{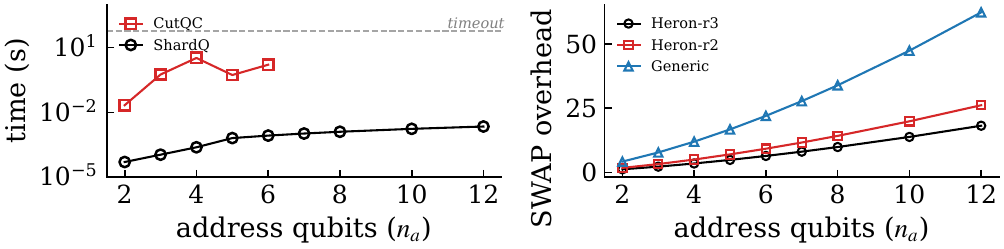}
    \caption{(a) Classical cut-finding time on \qcrank\ encoders. \sq\ sorts $\mathcal{G}_{\mathrm{cross}}$. CutQC and Qiskit Knitting are MIP searches and stop at the 60-second timeout. (b) Projected SWAP-overhead model, $\epsilon_2 \cdot 3 \cdot n_a^{1.5}$, on the coupling graphs named in the legend. Measured two-qubit counts at fixed $k$ are in \cref{tab:routing}.}
    \label{fig:cut_finding}
\end{figure}

CutQC and Qiskit Circuit Knitting search for a general circuit partition with a mixed-integer program, and we run both with a 60-second timeout. CutQC is now integrated into the Qiskit Circuit Knitting toolbox, so we treat the two as related MIP baselines \footnote{https://github.com/PrincetonUniversity/cutqc2?tab=readme-ov-file}. SparseCut only sorts the cross-register gates of an address--data encoder by coupling-graph distance, at cost $O(|\mathcal{G}_{\mathrm{cross}}|\log |\mathcal{G}_{\mathrm{cross}}|)$. On this circuit family those solvers reach the timeout beyond 6 address qubits (\cref{fig:cut_finding}(a)). This is a search-time comparison on encoders whose candidates are $\mathcal{G}_{\mathrm{cross}}$. Partition quality at fixed $k$ is \cref{tab:routing}.

\begin{figure}[t]
    \centering
    \includegraphics[width=0.74\linewidth]{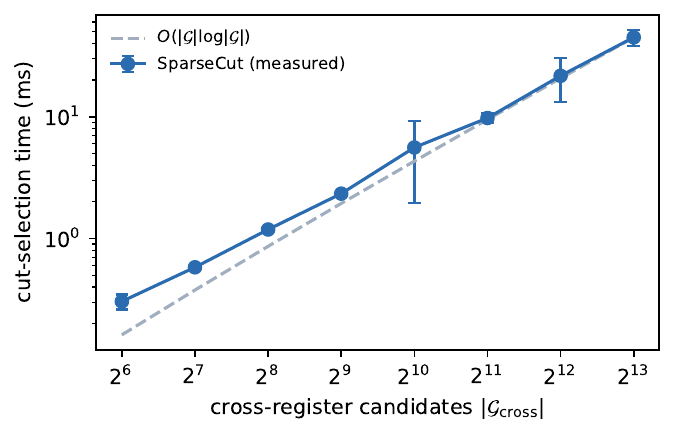}
    \caption{SparseCut cut-selection time versus the number of cross-register candidates $|\mathcal{G}_{\mathrm{cross}}|$ (mean $\pm$ std over 20 repeats). The measured time tracks the $O(|\mathcal{G}_{\mathrm{cross}}|\log|\mathcal{G}_{\mathrm{cross}}|)$ reference and stays under $50$~ms at device scale.}
    \label{fig:sparsecut_time}
\end{figure}

\paragraph{Cut-selection scaling}
\Cref{fig:sparsecut_time} isolates the cost of the SparseCut ranking itself on expanded \qcrank\ encoders ($n_d=8$), averaged over 20 repeats. Because SparseCut only enumerates and sorts the cross-register candidate set, its selection time follows the near-linearithmic $O(|\mathcal{G}_{\mathrm{cross}}|\log|\mathcal{G}_{\mathrm{cross}}|)$ complexity of \cref{alg:sparse_cut_selection}, remaining well under $50$~ms across the tested range and beyond the $156$-qubit device scale. This is consistent with the sub-second cut selection in \cref{fig:cut_finding}(a) and contrasts with the MIP-based baselines that reach the configured timeout for larger instances.

\begin{figure}[htbp]
    \centering
    \includegraphics[width=0.98\linewidth]{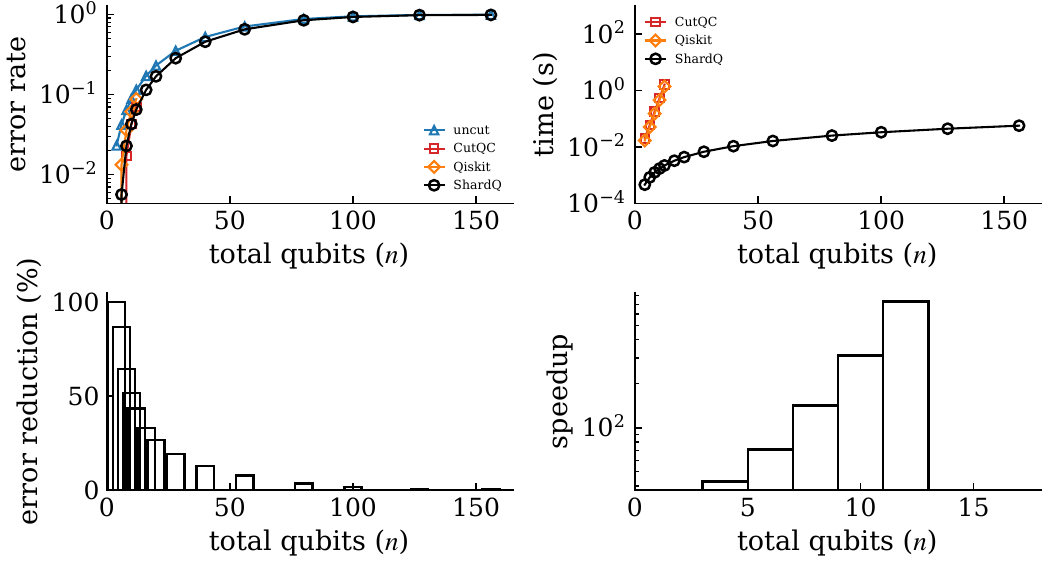}
    \caption{Projected cost model. (Top-Left) Modeled error rate. (Top-Right) Modeled cut-finding time, with MIP curves stopped at 60 seconds. (Bottom-Left) Modeled error reduction relative to the uncut model. (Bottom-Right) Modeled search-time ratio on instances where the MIP model finishes inside 60 seconds. The gray band is the projected large-$n$ region. Hardware RVF is reported in \cref{tab:noise}.}
    \label{fig:baseline_comparison}
\end{figure}

On \texttt{ibm\_pittsburgh}, \cref{tab:noise} is the hardware comparison with the uncut \qcrank\ encoder. At $n_a=4$, $n_d=8$, $k=2$, and $10^6$ shots on each submitted circuit, RVF is 0.79 for the uncut circuit and 0.90 for \sq. \Cref{fig:baseline_comparison} is the projected cost model named in its caption.
\Cref{tab:routing} and \cref{fig:cut_strategy} compare three selectors at the same cut count $k=2$: SparseCut ($d_T$), logical-distance ranking, and uniform random selection over $\mathcal{G}_{\mathrm{cross}}$. The study is a classical transpilation on a heavy-hex coupling map. Every selector reuses the initial layout of one uncut transpilation at optimization level 3 and transpiler seed 20240719, with $n_d=4$. The recorded quantity is the combined transpiled two-qubit count (CX, CZ, ECR, and SWAP). Random entries are the mean and standard deviation over 8 seeds. The calibration-weighted score \cref{eq:cal-score} is defined for the same candidate set and is not the ranking used in \cref{tab:routing}. At $n_a=3$ and $n_a=4$, logical-distance ranking removes a larger share of two-qubit gates than hardware distance. At $n_a=6$, hardware distance is the only one of the three selectors with a positive reduction. The routing benefit of $d_T$ is therefore specific to this encoder and to the address-register size.

\subsection{Analytical Cost Model}
\label{sec:costmodel}
We model the trade-off between the hardware benefit of removing a long-range gate and the classical cost of reconstructing it (RQ4). Let $\epsilon_2$ denote the median two-qubit (CZ) error, and let the $i$-th selected cross-register gate compile to $s_i$ SWAP operations under the layout $\phi$, each expanding to three CZ operations on heavy-hex hardware. Cutting the $k$ highest-scoring gates removes an accumulated routing error
\begin{equation}
\Delta_{\mathrm{hw}}(k) \;\approx\; \sum_{i=1}^{k} 3\, s_i\, \epsilon_2 ,
\label{eq:hwgain}
\end{equation}
where SparseCut orders candidates by hardware graph distance (\cref{eq:sparsecut-score}) so that $s_1 \ge s_2 \ge \cdots$. The reconstruction cost in this model is the sampling overhead $9^{k}$ of \cref{eq:overhead} \cite{mitarai2021constructing,peng2020simulating,pashayan2015estimating,brenner2023optimal,schmitt2025cutting}, and the variance of the knitted estimator grows as $\mathcal{O}(9^{k}/N)$ for a fixed total shot budget $N$. That variance factor is distinct from the six measurement settings per cut in \cref{tab:phi_prime_bases}. Since $\Delta_{\mathrm{hw}}(k)$ has diminishing increments while the sampling penalty grows geometrically, the net benefit in this model is largest at small $k$, consistent with the marked point $k=2$ in \cref{fig:ablation}. \sq\ is evaluated in that small-cut regime.

\subsection{Scalability and Overhead}
\label{sec:scalability}
\begin{figure}[t]
    \centering
    \begin{subfigure}[b]{0.49\linewidth}
        \centering\includegraphics[width=\linewidth]{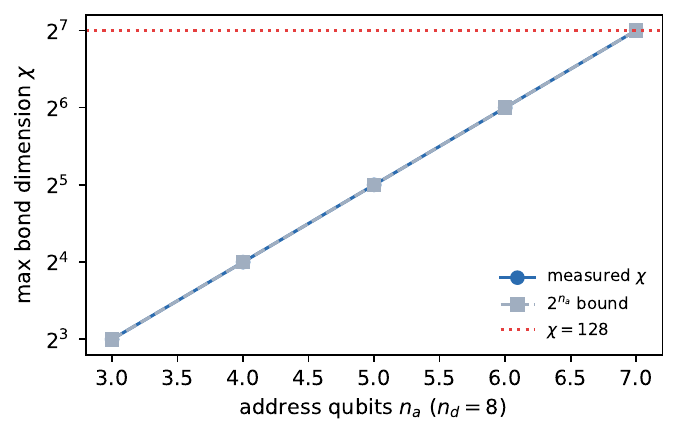}
        \caption{}\label{fig:mps_bonddim_a}
    \end{subfigure}
    \hfill
    \begin{subfigure}[b]{0.49\linewidth}
        \centering\includegraphics[width=\linewidth]{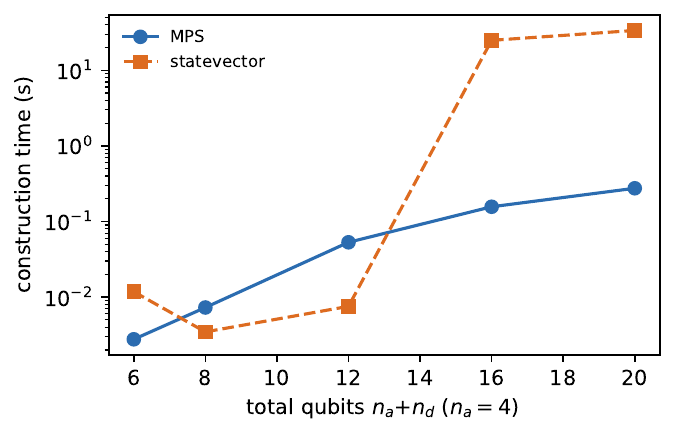}
        \caption{}\label{fig:mps_bonddim_b}
    \end{subfigure}
    \caption{Classical Aer MPS simulation of \qcrank\ encoders. (a) Exact bond dimension $\chi=2^{n_a}$ at $n_d=8$; $\chi\le 128$ for $n_a\le 7$. (b) Construction time at fixed $n_a=4$.}
    \label{fig:mps_bonddim}
\end{figure}

\begin{figure}[htbp]
    \centering
    \begin{subfigure}[b]{0.48\linewidth}
        \centering
        \includegraphics[width=\linewidth]{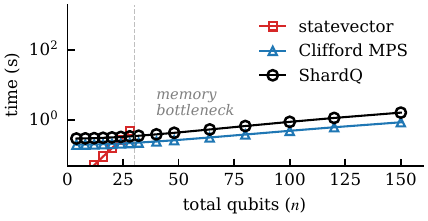}
        \caption{MPS Simulation Scaling}
        \label{fig:mps_scaling}
    \end{subfigure}
    \hfill %
    \begin{subfigure}[b]{0.48\linewidth}
        \centering
        \includegraphics[width=\linewidth]{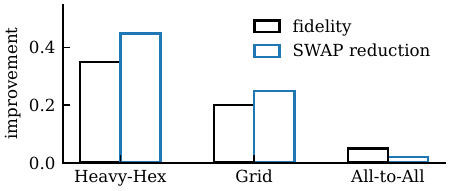}
        \caption{Topological Sensitivity}
        \label{fig:hardware_topology}
    \end{subfigure}

    \par\bigskip %

    \begin{subfigure}[b]{0.99\linewidth}
        \centering
        \includegraphics[width=\linewidth]{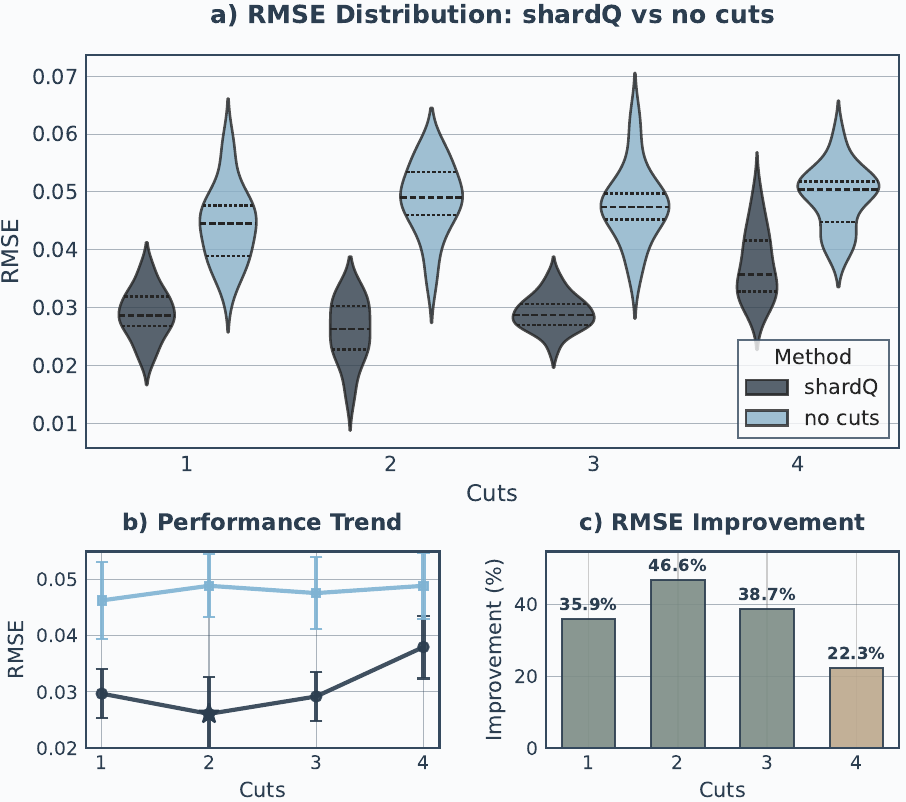}
        \caption{Cut Count Ablation}
        \label{fig:ablation}
    \end{subfigure}
    \hfill %
    \begin{subfigure}[b]{0.99\linewidth}
        \centering
        \includegraphics[width=\linewidth]{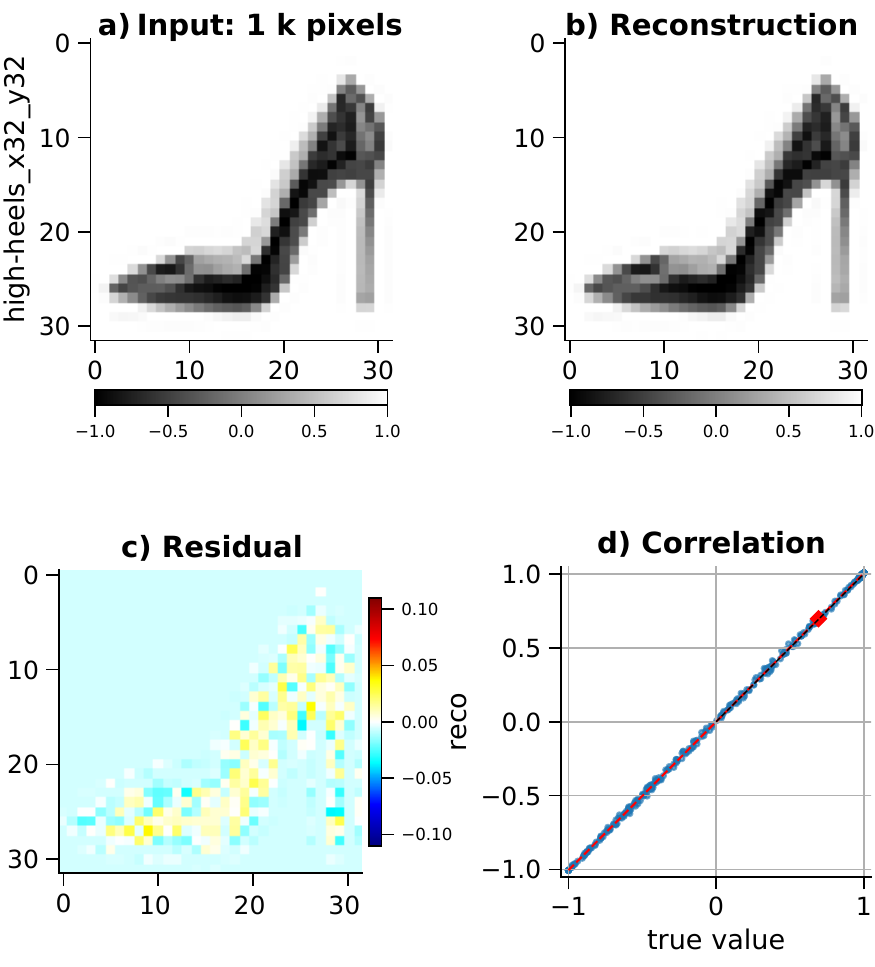}
        \caption{Image Reconstruction}
        \label{fig:res3}
    \end{subfigure}

    \caption{(a) Classical MPS timing. The portion past the sizes contracted with Aer is a projection. (b) Illustrative model of fidelity gain and SWAP reduction by topology. (c) Cut-count study. The marked operating point is $k=2$, and the sampling overhead is $9^k$ (\cref{eq:overhead}). (d) Noiseless classical reconstruction, RVF only (\cref{eq:rvf}). Hardware RVF is in \cref{tab:noise}.}
    \label{fig:unified_results}
\end{figure}

\begin{figure}[t]
    \centering
    \includegraphics[width=0.98\linewidth]{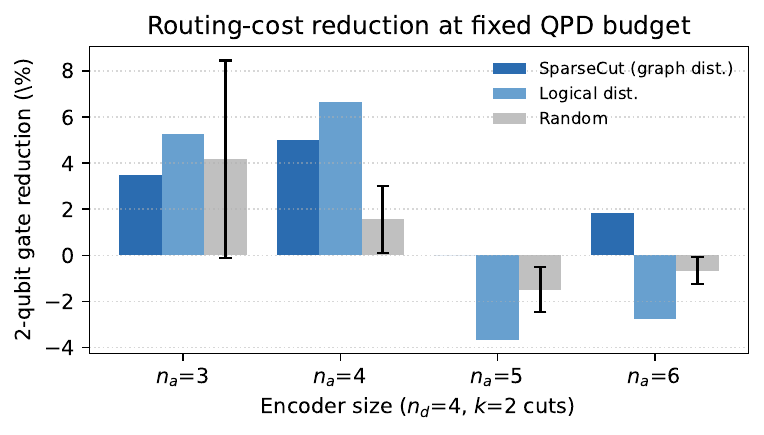}
    \caption{Classical transpilation study at fixed $k=2$ and $n_d=4$. Bars are the percent reduction in transpiled two-qubit count relative to the uncut circuit. Random bars show the mean over 8 seeds. Numerical values are in \cref{tab:routing}.}
    \label{fig:cut_strategy}
\end{figure}

Truncated subcircuits have reduced long-range entanglement because SparseCut removes the highest-scoring cross-partition gates. For these encoders the exact bond dimension is $\chi=2^{n_a}$, so $\chi \le 128$ on the window $n_a \le 7$ (\cref{fig:mps_bonddim}). \Cref{fig:mps_scaling} shows classical MPS timings. The 150-qubit point on that curve is a projection past the sizes contracted with Aer, and the caption marks the projected portion.

\paragraph{Bond-dimension boundedness}
We extend the scaling analysis by measuring the exact maximum bond dimension $\chi$ required to represent the encoder state, using the Aer matrix-product-state backend. \Cref{fig:mps_bonddim}(a) is a classical simulation of this \qcrank\ encoder with $n_d=8$. The exact maximum bond dimension is $\chi = 2^{n_a}$, so $\chi \le 128$ holds for $n_a \le 7$, and each added address qubit doubles $\chi$. At fixed $n_a=4$, construction time changes slowly with $n_d$ (\cref{fig:mps_bonddim}(b)). The memory $O(n\chi^2)$ therefore scales as $O(n\, 4^{n_a})$ for this encoder.

\Cref{fig:hardware_topology} is an illustrative model of fidelity gain and SWAP reduction on heavy-hex, grid, and all-to-all layouts. On the heavy-hex placement used for \cref{tab:routing}, address--data distances of the selected cuts are 4--6. Measured two-qubit counts before and after $k=2$ cuts are the columns of \cref{tab:routing}.
RVF and RMSE measure different objects. RVF is the squared Bhattacharyya coefficient between bit-string distributions, \cref{eq:rvf}. RMSE is computed after the decoder turns those probabilities into pixel angles through $\theta = 2\arctan\sqrt{|s|^2/|c|^2}$. \Cref{fig:res3} is a noiseless classical reconstruction with RVF above 0.99, which says the signed QPD sum reproduces the ideal distribution when shot noise is negligible. \Cref{tab:noise} is the hardware distribution comparison: on \texttt{ibm\_pittsburgh}, Experiment \#2, RVF is 0.79 for the uncut encoder and 0.90 for \sq, at $10^6$ shots on each submitted circuit. RMSE for those same shots is not stored with \cref{tab:noise}. The figure does not sweep either metric against the shot budget.
Cutting $k$ CX gates multiplies the sampling overhead by $9^k$ (\cref{eq:overhead}). The number of executed measurement settings is six per cut (\cref{tab:phi_prime_bases}). \Cref{fig:ablation} plots the cut-count study at the per-circuit shot count used for that panel, and the marked operating point is $k=2$. Running the measurement settings on several nodes leaves the factor $9^k$ unchanged.

We further validated the protocol on IBM QPUs. \Cref{tab:noise} shows that \sq\ outperforms the uncut baseline in the paired Pittsburgh runs, achieving an RVF of 0.90 vs. 0.79 for Experiment \#2. This indicates that, at the tested cut budget, the reduction in circuit depth and routing error offsets the QPD sampling overhead.

\begin{table}[t]
\centering
\caption{IBM Heron executions. Each submitted circuit receives $10^6$ shots. One job per column, so no repeat interval is attached. RVF is \cref{eq:rvf} after the clipping map \cref{eq:clip}.}
\label{tab:noise}
\small
\setlength{\tabcolsep}{5pt} %
\begin{tabularx}{\columnwidth}{@{}Xccccc@{}} %
\toprule
\textbf{Exp.} & \#1 & \#2 & \#3 & \#4 & \#5 \\
\textbf{Method} & \qcrank & \sq & \qcrank & \sq & \sq \\
\midrule
\textbf{Device} & \multicolumn{4}{c}{\textbf{ibm\_pittsburgh}} & \textbf{marrakesh} \\
\textit{CZ err} & \multicolumn{4}{c}{($1.46 \times 10^{-3}$)} & -- \\
\midrule
$n_a$ & 4 & 4 & 5--7 & 5--7 & 4 \\
$n_d$ & 8 & 8 & 8 & 8 & 8 \\
\# Addr. & 16 & 16 & 32--128 & 32--128 & 16 \\
Input (b) & 128 & 128 & 256--1k & 256--1k & 128 \\
\midrule
\textbf{RVF} & 0.79 & \textbf{0.90} & 0.75 & \textbf{0.88} & 0.75 \\
\bottomrule
\end{tabularx}
\end{table}

\begin{table}[t]
\centering
\caption{Classical routing comparison at fixed cut count $k=2$ and $n_d=4$ on a heavy-hex coupling map. All selectors share the initial layout of one uncut transpilation (Qiskit optimization level 3, transpiler seed 20240719). Counts are transpiled two-qubit gates (CX, CZ, ECR, and SWAP combined). Reduction is relative to the uncut count. Random is the mean and standard deviation of that reduction over 8 seeds. The sampling overhead is $9^{2}=81$ for every row. This table is not a QPU measurement and does not use \cref{eq:cal-score}.}
\label{tab:routing}
\small
\setlength{\tabcolsep}{3pt}
\begin{tabular}{@{}cccccc@{}}
\toprule
$n_a$ & $|\mathcal{G}_{\mathrm{cross}}|$ & Uncut & SparseCut & Logical & Random (\%) \\
\midrule
3 & 32 & 57 & 55 (3.5\%) & 54 (5.3\%) & $4.2\pm 4.3$ \\
4 & 64 & 120 & 114 (5.0\%) & 112 (6.7\%) & $1.6\pm 1.5$ \\
5 & 128 & 219 & 219 (0.0\%) & 227 ($-3.7\%$) & $-1.5\pm 1.0$ \\
6 & 256 & 432 & 424 (1.9\%) & 444 ($-2.8\%$) & $-0.7\pm 0.6$ \\
\bottomrule
\end{tabular}
\end{table}

\section{Conclusion}
\label{sec:disc}
In this study, we present \sq, a hardware-aware cutting and recomposition framework for structured quantum data encoding circuits. SparseCut formulates cut selection as a deterministic ranking over the cross-register candidate set $\mathcal{G}_{\mathrm{cross}}$, giving a cut-selection cost of $O(|\mathcal{G}_{\mathrm{cross}}|\log |\mathcal{G}_{\mathrm{cross}}|)$ after candidate enumeration. On the address--data encoders in this paper, that sort is the search-time comparison in \cref{fig:cut_finding}. The MIP baselines reach the 60-second timeout beyond 6 address qubits. Fixed-$k$ routing quality, including random selection, is \cref{tab:routing}.

For the CX decomposition used here, the sampling overhead is $9^k$ \cite{mitarai2021constructing,peng2020simulating,pashayan2015estimating,brenner2023optimal,schmitt2025cutting}, the signed sum in \cref{eq: T1} contains $8^{M}$ completions, each cut is executed as the six measurement settings in \cref{tab:phi_prime_bases}, and the variance at a fixed total of $N$ shots scales as $\mathcal{O}(9^{k}/N)$. \Cref{tab:noise} is the hardware result at $k=2$ with $10^6$ shots on each submitted circuit. Those entries are single jobs, so the repeat interval $\bar{x}\pm 1.96\,s/\sqrt{R}$ is not attached. The histogram is the clipped distribution \cref{eq:clip}. The MPS bond dimension measured for these encoders is $2^{n_a}$ (\cref{fig:mps_bonddim}). Large-$n$ curves that extend past the contracted sizes are marked as projections in their captions.

\section*{Acknowledgements}
This research used resources of the National Energy Research
Scientific Computing Center, a DOE Office of Science User Facility
supported by the Office of Science of the U.S. Department of Energy
under Contract No. DE-AC02-05CH11231 using NERSC award
NERSC DDR-ERCAP0034486.

\bibliographystyle{IEEEtran}
\bibliography{main}

\appendices
\section{Theoretical Analysis of Quasi-Probability Decomposition}
\label{app:qpd}
In this section, we present a detailed proof of the decomposition of the \cnot\ gate within our protocol, which involves two address qubits and one data-qubit tensor data encoder.
Notably, the permutation of $\UCRy$ gates represents a recent approach for differential data encoding blocks, which can be realized through the constant-depth quantum circuits demonstrated in \cite{faro2024families}.
Specifically, the single rotation gate and controlled Y rotation gate are
\begin{equation}
\begin{aligned}
R_y(\theta) &= \cos\left(\frac{\theta}{2}\right) I - i\sin\left(\frac{\theta}{2}\right) Y, \\
CRY(\theta) &= |0\rangle\langle 0| \otimes I + |1\rangle\langle 1| \otimes R_y(\theta).
\end{aligned}
\end{equation}
The commutative principle is expressed in Eq. (18-21) of \cite{jan1}, where the $\CRY$ can be decomposed into a $\Ry$ with a \cnot\ gate. Such a decomposition enables the construction of an encoder circuit with two address qubits and one data qubit, denoted as $\mathcal{C}_{21}$ shown in \cref{circ: 21}.
\begin{figure*}[htbp]
    \centering
    \begin{quantikz}[row sep=0.25em,column sep=0.75em]
    \lstick{$a_0$} & \gate{H}   &                                      &          &                                      & \ctrl{2}   &      &       &   & \ctrl{2}  &\meter{}\\
    \lstick{$a_1$} & \gate{H}   &                                      & \ctrl{1} &                                      &            &   & \ctrl{1} &  &   &\meter{}\\
    \lstick{$d^\star$} &        & \gate{R_y(P_0\lbrace a_{10}\rbrace)} & \targ{}  & \gate{R_y(P_0\lbrace a_{11}\rbrace)} & \targ{}    & \gate{R_y(P_0 \lbrace a_{10}\rbrace)}  & \targ{}    & \gate{R_y(P_0\lbrace a_{00}\rbrace)} & \targ{}  &\meter{} \\
    \end{quantikz}
    \caption{The example of two address and one data qubits \qcrank.}
    \label{circ: 21}
\end{figure*}
Note that the permutation of \cnot\ is encoded with the gray code \cite{bookstein2002generalized} that optimizes the Hamming distance between the neighbors to a maximum value of one. The advantages become more apparent when we have more address qubits because the traversal of control configurations allows more efficient updating of \cnot\ control patterns; only one control bit changes between consecutive operations, which minimizes the number of \cnot\ gates required to reconfigure the multi-controlled rotation. We recall that the number of the address qubits is \(n_{a}\), hence, the $\Ry$ gates are parameterized by
\(P_0\{a_{00}\},\dots,\ P_0\{binary(2^{n_{a}}\}\).
Specifically, the data-to-angle encoding is given by

\begin{equation}
   \theta^{\mathrm{final}} = \mathrm{Gray} \left( \mathrm{FWHT} \left( \arccos(\mathbf{d}) \right) \right)
   \label{eq: angle}
\end{equation}
where \( \mathbf{d} \in [-1, 1]^N \) is the input data vector, \( \arccos(\cdot) \) maps data to angles, \( \mathrm{FWHT} \) is the scaled fast Walsh-Hadamard transform, and the gray code permutation is used to optimize the control pattern for efficient quantum circuit simulation.

Given that the simplest maximum cut number set is one using \cref{alg:sparse_cut_selection}, we obtain the cutting gate index as the sixth from circuit $\mathcal{C}_{21}$.
Then, considering that the uncut circuit is decomposed into six subcircuits because of the Pauli group, each subcircuit is attained through a probabilistic representation of the Clifford gate.
Here, the KAK decomposition for $\mathcal{C}_{21}$ is given by implementing a uniformly controlled rotation with address qubits $q_0, q_1$ and data qubit $q_2$ denoted by
\begin{equation}
|\psi\rangle = \frac{1}{2}\sum_{i=0}^{3} |i\rangle \otimes \left(\cos\frac{\alpha_i}{2}|0\rangle + \sin\frac{\alpha_i}{2}|1\rangle\right).
\label{eq: phi}
\end{equation}
Then, by recalling that \cref{sec:globalreco}, each controlled rotation gate can be decomposed to
\begin{equation}
\text{UCR}_y(\vec{\alpha}) = \sum_{a_1,a_2 \in \{\pm 1\}} c_{a_1,a_2} \cdot P_{a_1}^{(0)} \otimes P_{a_2}^{(1)} \otimes R_y(\theta_{a_1,a_2}).
\label{eq: UCRy}
\end{equation}
where $P$ is the Pauli measurement projector. We note that the number of cut indices results in the $\mathcal{O}(n)$ linearly increasing of the QPD measurement stored as the temporary result with the coefficient shown in \cref{tab: pauli}. This is because \textbf{SparseCut} allows cuts at the data block level, as shown in \cref{fig:method}.
To prove the \cnot\ cut, we first recall the CZ gate decomposition shown in Fig. 6 of \cite{peng2020simulating}. Additionally, the \cnot\ can be represented by $CZ$ with
\begin{equation}
   \cnot\;=\;(I\!\otimes\! H)\,\text{CZ}\,(I\!\otimes\! H).
   \label{eq:CX}
\end{equation}
However, only three terms (Pauli X, Pauli Z, and Hadamard gate) are required for the \qcrank. Here, $\CZ$ can be summed by $\Rz$ %
because of the two virtual qubit gate decomposition principle \cite{mitarai2019methodology}
\begin{equation}
    \text{CZ} = \sum_{a_1,a_2 \in \{\pm 1\}^2} a_1 a_2 \left\{ RZ\left(\frac{a_1\pi}{2}\right) \otimes RZ\left(\frac{a_2\pi}{2}\right) \right\}.
    \label{eq:CZdecom}
\end{equation}
With \cref{eq:CZdecom}, it applies the conjugation to \cref{eq:CX}
\begin{align}
\cnot\ =\ & (I \otimes H) \cdot \Bigg[ \sum_{a_1,a_2 \in \{\pm 1\}^2} a_1 a_2 \Big\{ \notag\\
& \qquad RZ\left(\frac{a_1\pi}{2}\right) \otimes RZ\left(\frac{a_2\pi}{2}\right) \Big\} \Bigg] \cdot (I \otimes H)
\end{align}
Because $H \cdot RZ(\theta) \cdot H = RX(\theta)$ (referred to Clifford decomposition table \cref{tab:gate_decomposition}), this gives us separate terms conditioned based on the coefficients
\begin{equation}
    H \cdot RZ\left(\frac{a_2\pi}{2}\right) \cdot H = RX\left(\frac{a_2\pi}{2}\right).
\end{equation}
The general decomposition for the two virtual qubit gates can be written in the format of superoperators with a single qubit operation sandwiching the observable density matrix. Let us expand the observables into three Pauli matrices and identity term, and the coefficients are defined by~\cref{tab: pauli}
\begin{table}[htbp]
\centering
\small
\setlength{\tabcolsep}{3pt}
\renewcommand{\arraystretch}{1.05}

\begin{tabular*}{\columnwidth}{@{\extracolsep{\fill}}
c c c c | c c c c | c c c c
}
\toprule
$i$ & $O_i$ & $\rho_i$ & $c_i$ &
$i$ & $O_i$ & $\rho_i$ & $c_i$ &
$i$ & $O_i$ & $\rho_i$ & $c_i$ \\
\midrule
1 & $I$ & $\ket{+}\!\bra{+}$ & $+\tfrac12$ &
4 & $X$ & $\ket{1}\!\bra{1}$ & $-\tfrac12$ &
7 & $Z$ & $\ket{+}\!\bra{+}$ & $+\tfrac12$ \\

2 & $I$ & $\ket{-}\!\bra{-}$ & $+\tfrac12$ &
5 & $Y$ & $\ket{-i}\!\bra{-i}$ & $+\tfrac12$ &
8 & $Z$ & $\ket{-}\!\bra{-}$ & $-\tfrac12$ \\

3 & $X$ & $\ket{0}\!\bra{0}$ & $+\tfrac12$ &
6 & $Y$ & $\ket{+i}\!\bra{+i}$ & $-\tfrac12$ &
 &  &  &  \\
\bottomrule
\end{tabular*}

\caption{Parameters defining the operator $O_i$, state $\rho_i$, and coefficient $c_i$ for each index $i$.}
\label{tab: pauli}
\end{table}

\newcommand{\gdot}{\,\cdot\,\allowbreak} %

\begin{table}[htbp]
\centering
\scriptsize
\setlength{\tabcolsep}{3pt}
\renewcommand{\arraystretch}{1.05}

\newcommand{\DecW}{0.23\columnwidth}

\begin{tabular*}{\columnwidth}{@{\extracolsep{\fill}}l|p{\DecW} l|p{\DecW} l|p{\DecW}}
\toprule
\textbf{Gate} & \textbf{Decomposition} &
\textbf{Gate} & \textbf{Decomposition} &
\textbf{Gate} & \textbf{Decomposition} \\
\midrule
$X$        & $H\gdot Z\gdot H$ &
$R_Z$      & $S^\dagger$ &
$\Pi Y$    & $H\gdot S^\dagger\gdot H\gdot P_0\gdot H\gdot S\gdot H$ \\

$Y$        & $H\gdot Z\gdot H\gdot Z$ &
$R_{YZ}$   & $H\gdot S^\dagger\gdot H\gdot Z$ &
$\Pi Z$    & $P_0$ \\

$Z$        & $S^2$ &
$R_{ZX}$   & $S^\dagger\gdot H\gdot S^\dagger\gdot H\gdot S^\dagger$ &
$\Pi_{YZ}$ & $S\gdot H\gdot S\gdot H\gdot P_0\gdot H\gdot S\gdot H\gdot S^\dagger$ \\

$R_X$      & $H\gdot S^\dagger\gdot H$ &
$R_{XY}$   & $H\gdot Z\gdot H\gdot S^\dagger$ &
$\Pi_{ZX}$ & $H\gdot S^\dagger\gdot H\gdot P_0\gdot H\gdot S\gdot H\gdot Z$ \\

$R_Y$      & $S\gdot H\gdot S^\dagger\gdot H\gdot S^\dagger$ &
$\Pi X$    & $S\gdot H\gdot S\gdot H\gdot P_0\gdot H\gdot S^\dagger\gdot H\gdot S^\dagger$ &
$\Pi_{XY}$ & $P_0\gdot H\gdot Z\gdot H$ \\
\bottomrule
\end{tabular*}

\caption{Decomposition of gates in terms of $H$ Hadamard gate, $S$ Phase gate, $S^\dagger$ Reverse phase gate, $Z$ Pauli Z gate, and $P_0$ Z-based measurement gate used in controlled-rotation circuits. We note that the decomposition for the Pauli groups is defined in the Clifford gate group \cite{bravyi2005universal} except for the Z measurement gate. In local operator classic communication, \qpd\ stores the mid-circuit measurement result for the global %
reconstruction.}
\label{tab:gate_decomposition}
\end{table}

\begin{figure}[htbp]
\centering
\resizebox{\columnwidth}{!}{%
\begin{tikzpicture}[scale=0.9, thick, >=latex,
    gate/.style  = {draw, fill=white, text height = 1.5ex, text depth = .25ex, minimum width = 2.5em, minimum height = 2em},
    sgate/.style = {fill = white, draw, text height = 1.5ex, text depth = .25ex, minimum size = 1.5em, inner sep=2pt},
    dy/.style = {yshift=0.5cm},
    ndy/.style = {yshift=-0.5cm}
]

\node at (0, 0.5) {$q_c$};
\node at (0, -0.5) {$q_t$};
\draw[->] (0.3, 0.5) -- (2.0, 0.5);
\draw[->] (0.3, -0.5) -- (2.0, -0.5);

\node (CX) [gate] at (1.15, 0) {$\text{CNOT}$};

\node at (2.4, 0) {$=$};

\draw[->] (2.8, 0.5) -- (3.8, 0.5);
\draw[->] (2.8, -0.5) -- (3.8, -0.5);

\node at (4.1, 0) {$+$};

\draw[->] (4.4, 0.5) -- (5.6, 0.5);
\draw[->] (4.4, -0.5) -- (5.6, -0.5);
\node[sgate] at (5, 0.5) {$Z$};
\node[sgate] at (5, -0.5) {$X$};

\node at (5.9, 0) {$+$};

\node at (7.0, -0.05) {$\displaystyle \sum_{a_1,a_2} a_1 a_2$};

\draw[->] (8.2, 0.5) -- (10.5, 0.5);
\draw[->] (8.2, -0.5) -- (10.5, -0.5);

\node[sgate] at (9.35, 0.5) {\small $\frac{I + a_2 Z}{2}$};
\node[sgate] at (9.35, -0.5) {\small $H S^{a_1} H$};

\draw[decorate,decoration={brace,amplitude=1.5ex},xshift=2pt]
  (8.2, -0.8) -- (8.2, 0.8);

\node at (10.8, 0) {$+$};

\draw[->] (11.1, 0.5) -- (13.4, 0.5);
\draw[->] (11.1, -0.5) -- (13.4, -0.5);

\node[sgate] at (12.25, 0.5) {\small $H S^{a_1} H$};
\node[sgate] at (12.25, -0.5) {\small $\frac{I + a_2 X}{2}$};

\draw[decorate,decoration={brace,amplitude=1.5ex,mirror},xshift=-2pt]
  (13.4, -0.8) -- (13.4, 0.8);

\end{tikzpicture}%
}
\caption{CX-decomposition using only Clifford gates ($H, S, Z, X$) and projectors.
Rotations $R_Z(\pm\pi/2)$ are replaced by $S/S^\dagger$, and $R_X(\pm\pi/2)$ by $H S H$.}
\label{fig:CX-decomp-clifford}
\end{figure}
Here, $\mathcal{O}_i$ is the observable, $\rho_i$ is the eigenstate, $c_i$ is the coefficient. Note that the cut gate has one prepared state and measures an observable. Specifically, on the preparation side, we apply a 1-qubit density matrix $\rho_\lambda = |\lambda\rangle\!\langle\lambda|$ that is the eigenstate of the Pauli as appearing in \cref{tab: pauli}. On the measurement side, $s=\pm1$ serves as the measured and recorded eigenvalues. In this experiment, we used gate-based quantum circuit simulations. After shot-averaging,
\begin{equation}
   \bigl\langle s\bigr\rangle
=\operatorname{Tr}\bigl[ P\,\rho_\lambda \bigr]
=\lambda ,
\label{eq: qpd_meas}
\end{equation}
so the product of eigenvalue and shot average reproduces the Pauli and we denote the QPD Pauli representation of \cnot\ gate as
\begin{equation}
    {\rm \cnot}_{c\to\ \!t}
=\frac12\Bigl(
      I_c\!\otimes\!I_t
    + Z_c\!\otimes\!I_t
    + I_c\!\otimes\!X_t
    - Z_c\!\otimes\!X_t
\Bigr),
\label{eq: CNOT_all}
\end{equation}
where c is the conditioned (control) qubit and t is the target qubit.
Therefore, to generalize the eight Pauli observable and re-prepare for the global states, we categorize the six terms denoted in \cref{tab:phi_prime_bases}.
\begin{table}[htbp]
  \centering
  \footnotesize
  \caption{Post-measurement bases associated with the six transformed
           Pauli–Stinespring operators \(\Phi'_i\).
           $S$: unitary on both sides, $M_A$: measure \& reprepare qubit 1, $M_B$: mid-measurement on qubit 2.}
  \label{tab:phi_prime_bases}
  \begin{tabular}{c|c|c}
    \toprule
    $\Phi'_i$ from (E1~\cite{mitarai2021constructing}) & Applied gates & Computation basis \\
    \midrule
    $S(I\otimes I)$              & no mid measurement & identity term \\
    $S(A\otimes B)$              & no mid measurement & $Z\otimes X$ term \\
    $M_A\otimes S(e^{i\pi B/4})$ & $S^\dagger H$ + $Z$-meas. & $S^\dagger$ basis ($|\pm i\rangle$) \\
    $M_A\otimes S(e^{-i\pi B/4})$& $SH$ + $Z$-meas. & $S$ basis ($|\mp i\rangle$) \\
    $S(e^{i\pi A/4})\otimes M_B$ & $I$ + $Z$-meas. & computational ($|0\rangle,|1\rangle$) \\
    $S(e^{-i\pi A/4})\otimes M_B$& $H$ + $Z$-meas. & Hadamard ($|\pm\rangle$) \\
    \bottomrule
  \end{tabular}
\end{table}
Here, \cref{alg:sparse_cut_selection} selects the Clifford bases for efficient quantum hardware simulation because of the efficiency of the overhead simulation of the Clifford group in this case.
We note that rows 1 and 2 share the same hardware setting; the difference is an $S$ versus\ $S^{\dagger}$ gate; likewise, 3 and 4 share Hadamard and no Hadamard settings. Each basis yields two possible classical outcomes, which correspond to the “\(\pm\)” states listed in the table.
Because distinct quantum circuits (measurement settings) are required, the eight table rows are recovered by the classical postprocessing of the outcomes from the six circuits. Therefore, to produce the single cut shown in \cref{circ: 21}, we combine \cref{eq: CNOT_all} and \cref{tab: pauli}. Hereby we proved \cref{fig:CX-decomp-clifford}.
Note that in the transpiled version of quantum circuit mapping to physical qubits, we do not consider the $\UCRy$ gate in the \textbf{SparseCut} algorithm.
To complete the proof for the generalized $\UCRy$, it is necessary to combine \cref{eq: UCRy} with E1 \cite{mitarai2021constructing}, resulting in the complete decomposition of the $\UCRy$ gate
\begin{align}
&\text{UCR}_y(\vec{\alpha}) = \frac{1}{4}\sum_{j=1}^{6} w_j \cdot \mathcal{M}_j \otimes \mathcal{R}_j \\
&= \frac{1}{4}\left[
\begin{array}{l}
w_1 \langle +i| \otimes R_y(\theta_1) + w_2 \langle -i| \otimes R_y(\theta_2) \\
+ w_3 \langle +i| \otimes I \cdot R_y(\theta_3) + w_4 \langle -i| \otimes Z \cdot R_y(\theta_4) \\
+ w_5 \langle 0| \otimes R_y(\theta_5) + w_6 \langle +| \otimes R_y(\theta_6)
\end{array}
\right].
\end{align}
For each subcircuit produces measurement outcomes with probabilities
\begin{equation}
P(m_j = \pm 1) = \frac{1 \pm \langle \psi | \sigma_j^{(0)} \otimes I^{(1)} \otimes I^{(2)} | \psi \rangle}{2}
\end{equation}
where $\ket\phi$ is defined in \cref{eq: phi}.
Finally, we produce the global measurement reconstruction using \cref{alg:recon}.
The original expectation values are recovered through
\begin{equation}
\langle R_y(\alpha_i) \rangle = \sum_{j=1}^{6} c_j \cdot \text{Corr}(m_j^{(0)}, m_j^{(2)})
\end{equation}
where $\text{Corr}(m_j^{(0)}, m_j^{(2)})$ is the correlation between the address and data qubit measurements in subcircuit $j$,  completing the end-to-end quantum data encoder circuit partitioning and recomposition process.

\end{document}